\documentclass[pre,twocolumn,aps,showpacs]{revtex4-1} 

\usepackage{amsmath} 
\usepackage{amssymb} 
\usepackage{graphicx}
\usepackage{color}

\begin{document}

\title{Time asymmetry of the Kramers equation with nonlinear friction: \\
fluctuation-dissipation relation and ratchet effect}

\author{A. Sarracino}
\affiliation{CNR-ISC and Dipartimento di Fisica, Universit\`a Sapienza, p.le A. Moro 2, 00185 Roma, Italy}

\begin{abstract}
We show by numerical simulations that the presence of nonlinear
velocity-dependent friction forces can induce a finite net drift in
the stochastic motion of a particle in contact with an equilibrium
thermal bath and in an asymmetric periodic spatial potential. In
particular, we study the Kramers equation for a particle subjected to
Coulomb friction, namely a constant force acting in the direction
opposite to the particle's velocity. We characterize the
nonequilibrium irreversible dynamics by studying the generalized
fluctuation-dissipation relation for this ratchet model driven by
Coulomb friction.
\end{abstract}


\maketitle

\section{Introduction}

Nonequilibrium systems show a rich phenomenology which is induced by
the presence of currents, breaking the time-reversal symmetry of the
dynamics. A measure of the lacking of detailed balance in systems
described within the framework of stochastic dynamics is given by the
entropy production functional~\cite{LS99,CM99}. Two main consequences
of the breaking of the time-reversal symmetry are the violations of
the equilibrium fluctuation-dissipation relation (EFDR),
see~\cite{CR03,BPRV08,seifert,BM13} for recent reviews, and the ratchet
effect~\cite{f63,R02}, namely the rectification of unbiased
fluctuations, when a spatial asymmetry is also introduced. Ratchet
systems (or Brownian motors) can be used to study important processes
of energy conversion to mechanical work and play a central role in
many different fields, such as energy harvesting from ambient
vibration~\cite{gamma} and transport in biological
systems~\cite{EWO98}.  Indeed, nature possesses an excellent command
of the subtle processes underlying such a phenomenon, as shown in the
cellular world: sophisticated mechanisms can realize the required
conversion of chemical energy into mechanical one, allowing
unidirectional motion, for instance of proteins and other
macro-molecules~\cite{SW03}.

At the theoretical level, a widely studied class of nonequilibrium
systems is represented by the Langevin equation describing the motion
of a particle in the presence of external forces and asymmetric
potentials. In this framework, a huge variety of ratchet models have
been studied in the last two decades (for reviews see~\cite{A97,JAP97,R02,HM03}).
Usually, in these studies external energy sources, such as
time-dependent oscillatory forces or time-dependent noise amplitudes,
are exploited in order to drive the system out of equilibrium and
break the time-reversal symmetry. For these systems, interesting
issues are, for instance, the energetics~\cite{PDC02,sekimoto,KG13} 
and efficiency at maximum power~\cite{SS07,T08,EKLB10,TBV11,GIP12,seifert},
the fluctuation theorems~\cite{LM09} and the role of motors coupling~\cite{BJP02,AMK08}. 

Recently the presence of nonlinear friction, in the form of Coulomb
(or dry) friction, modeled as a constant force opposite to the
particle's velocity, has been considered as another effective source
of dissipation.  The effect of the dry friction has been studied in
the context of the Langevin equation by de Gennes~\cite{dGen05}, who
showed that the diffusive properties of a Brownian particle are
strongly modified by the presence of Coulomb friction. In other
studies some interesting features of the Langevin equation with
nonlinear velocity-dependent friction have been put in evidence, as
for instance in~\cite{CD61,H05,DCdG05,PF07,TSJ10,BTC11,TPJ12,GC12}.  Within
the framework of ratchet systems the action of dry friction, coupled
with other dissipative forces, has been studied
in~\cite{FKPU07,talbot2,BS12}. Remarkably, it has been shown that such
a source of dissipation, which is usually deemed an hindrance to the
motion, can be sufficient to drive a ratchet effect, even if the
system is in contact with an equilibrium thermal
bath~\cite{gnoli,SGP13} and other external forces are absent.  In the
context of granular ratchets~\cite{cleuren2,CPB09,EWLM10,BV13}, recent
experiments have also shown the importance of the effect of dry
friction affecting the dynamics of tracers~\cite{GPT13,GSPP13}.

Here we bring to the fore the basic elements for a ratchet effect
driven by nonlinear velocity-dependent friction, analyzing a more
elementary system, with respect to the kinetic models studied
in~\cite{gnoli,SGP13}, where a master equation approach was
considered.  In particular, we show that the ratchet effect can be
obtained by considering a Langevin equation where a particle is in
contact with a thermal bath at a constant temperature and moves in an
asymmetric spatial potential, subjected to the action of nonlinear
velocity-dependent friction. No external force is considered, but the
unbiased thermal white noise of the classical Langevin equation.  We
first present a general proof that nonlinear velocity-dependent
friction force in a spatial potential leads to nonequilibrium
behaviors.  We then study numerically the Kramers equation, describing
the coupled evolution of position and velocity of the particle, and we
show that the source of dissipation introduced by the presence of
Coulomb friction is sufficient to induce nonequilibrium
conditions. These produce a violation of the EFDR and, due to the
asymmetric potential, a ratchet effect. This is a novel finding in the
context of Kramers equation. We also report similar results observed
in a model for active particles~\cite{SET98}, where the nonequilibrium
dynamics is generated by the presence of nonlinear velocity-dependent
friction coupled with an energy pumping term.

The paper is organized as follows.  In Sec.~\ref{1} we introduce the
Kramers equation with nonlinear friction and discuss the general
conditions to induce nonequilibrium currents. Then, in Sec.~\ref{FDT}
we derive a generalized fluctuation-dissipation relation (FDR) in
order to assess the nonequilibrium behavior of the system, pointing
out the violation of time-reversal symmetry. In Sec.~\ref{rat} we show
the ratchet effect for our model, and characterize its behavior for
two different forms of the frictional force.  Finally, in
Sec.~\ref{concl}, some conclusions are drawn.

\section{Kramers equation with nonlinear velocity-dependent friction force}
\label{1}

We consider the generalized Kramers equation for the motion of a
particle of mass $m=1$, with position $x$ and velocity $v$, 
in an external potential $U(x)$ in the presence
of a generic odd velocity-dependent friction term $F(v)$~\cite{K94}
\begin{eqnarray}\label{kramers}
\dot{x}(t) &=& v(t) \nonumber \\
\dot{v}(t)&=&-F[v(t)]-U'[x(t)] +\eta(t) +h(t), 
\end{eqnarray}
where $\eta(t)$ is a white noise, with $\langle \eta(t)\rangle=0$ and
$\langle \eta(t)\eta(t')\rangle=2\gamma T\delta(t-t')$, $\gamma$ and
$T$ being two parameters and $\delta(t)$ the Dirac's delta, and $h(t)$
is a small external perturbation (only used for measuring linear
response). 

\subsection{Detailed balance and nonequilibrium currents}
\label{DB}

In order to explicitly discuss the appearance of nonequilibrium
currents when nonlinear frictional forces are considered, let us write
the Fokker-Planck equation for the evolution of the density probability function
$P(x,v)$ of the process described by Eq.~(\ref{kramers})
\begin{eqnarray}
\frac{\partial P(x,v)}{\partial t}&=&-\frac{\partial}{\partial x}[vP(x,v)] 
+\frac{\partial}{\partial v}\{[F(v)+U'(x)]P(x,v)\} \nonumber \\
&+& \gamma T \frac{\partial^2}{\partial v^2}P(x,v)
= -\boldsymbol{\nabla}\cdot\boldsymbol{J}^{rev} - \boldsymbol{\nabla}\cdot\boldsymbol{J}^{irr}, 
\end{eqnarray}
where $\boldsymbol{\nabla}=(\partial_x,\partial_v)$ and we have
introduced the reversible and irreversible probability currents~\cite{R89},
respectively,
\begin{displaymath}
\boldsymbol{J}^{rev} =
\left( \begin{array}{c}
vP(x,v)  \\
-U'P(x,v) 
\end{array} \right)
\end{displaymath}
and
\begin{displaymath}
\boldsymbol{J}^{irr} =
\left( \begin{array}{c}
0  \\
-F(v)P(x,v)-\gamma T\partial_v P(x,v) 
\end{array} \right).
\end{displaymath}
These currents are obtained by decomposing the Fokker-Planck
operator under time reversal~\cite{R89} into a reversible (or streaming) operator 
and an irreversible (or collision) operator. 
The stationary condition reads
\begin{equation}
\boldsymbol{\nabla}\cdot(\boldsymbol{J}^{rev}+\boldsymbol{J}^{irr})=0.
\end{equation}
The system is said to be at equilibrium if the detailed balance
condition $\boldsymbol{J}^{irr}=0$ holds~\cite{R89} and this can only
happen in two situations: i) if the external spatial potential is
zero, namely $U=0$, or ii) if the frictional force is linear, namely
$F(v)=\gamma v$.  Such a result can be shown as follows.  From the equilibrium
condition $\boldsymbol{J}^{irr}=0$ one obtains the relation
\begin{equation}
\partial_vP(x,v)  =-\frac{F(v)}{\gamma T}P(x,v).
\label{cond}
\end{equation}
Then, using this relation in the stationary condition
$\boldsymbol{\nabla}\cdot\boldsymbol{J}^{rev}=0$, one obtains the
following form for the equilibrium distribution
\begin{equation}
P(x,v)=p(v)e^{-\frac{U(x)}{\gamma T}\frac{F(v)}{v}},
\label{eqform}
\end{equation}
where $p(v)$ is an unknown function which only depends on $v$. Now,
taking the derivative with respect to $v$ in Eq.~(\ref{eqform}) and
imposing the condition~(\ref{cond}) one has the following constraint
for $p(v)$
\begin{equation}
\frac{p'(v)}{p(v)}=\frac{U(x)}{\gamma T}\frac{F'(v)v-F(v)}{v^2}-\frac{F(v)}{\gamma T}.
\label{constr}
\end{equation}
Since in the left hand side of the above equation does not appear any
$x$-dependence, this constraint can be fulfilled only if $U(x)=0$ or
if $F'(v)v-F(v)=0$, namely if $F(v)\propto v$.  In~\cite{DHG09} it is
discussed how to introduce a multiplicative noise in order to recover
detailed balance with nonlinear velocity-dependent forces. See
also~\cite{Po13} for the case of diffusion in nonuniform temperature.

\subsection{Frictional forces and ratchet potential}

As the above discussion shows, for simple linear
friction $F[v(t)]=\gamma v(t)$, where here $\gamma$ is a viscous
friction coefficient, and in the absence of external force ($h=0$), the
system is characterized by the equilibrium stationary state
$P(x,v)\propto \exp\{-[v^2/(2T)+U(x)/T]\}$ at the temperature $T$.
The motion of a Brownian particle in a potential with linear
friction is widely studied in the literature, see for
instance~\cite{R89} and references therein.

As main instance of nonlinear velocity-dependent friction force, we
consider here also the presence of Coulomb friction, namely
\begin{equation}\label{coulomb}
F[v(t)]=\gamma v(t)+\Delta\sigma[v(t)], 
\end{equation}
where $\Delta$ is the frictional
coefficient and $\sigma(x)$ the sign function (with $\sigma(0)=0$). In the absence of external
potential, model~(\ref{kramers}) with frictional force~(\ref{coulomb}) 
has been studied for instance in~\cite{dGen05,H05,DCdG05,PF07,BTC11}.

Among other ways to introduce nonlinearity in the Kramers
equation, we also consider a model for active Brownian
particles~\cite{SET98}, inspired by the Rayleigh-Helmholtz model for
sustained sound waves~\cite{rayleigh}, where the frictional force is
\begin{equation}\label{active}
F[v(t)]=-\gamma_1 v(t)+\gamma_2 v^3(t),
\end{equation}
with $\gamma_1$ and $\gamma_2$ positive constants.
With this choice in Eq.~(\ref{kramers}) the motion of the particle is
accelerated for small $v$ and is damped for high $v$. This model
takes into account the internal energy conversion of the active
particles coupled to other energy sources.  In
the absence of external potential, the diffusion properties of a
Brownian particle with other forms of nonlinear frictional forces have
been studied also in~\cite{F04,PF07,BL08}.

In the following, in order to study unidirectional motion generated by
nonequilibrium fluctuations, we will consider the presence of an asymmetric ratchet
potential, focusing on the form usually studied in the literature
of Brownian ratchets~\cite{BHK94}, namely
\begin{equation}\label{ratchet_pot}
U(x)=\sin(x)+\mu\sin(2x), 
\end{equation}
where $\mu$ is an asymmetry parameter.
In the case of frictional force~(\ref{active}) the effect of an
asymmetric potential has been recently investigated in~\cite{STE00,FEGN08}.

\section{Fluctuation-dissipation relation}
\label{FDT}

The fluctuation-dissipation relation is one of the main achievement of
statistical mechanics. It allows one to express the response of a
system to an external perturbation in terms of spontaneous
fluctuations computed in the unperturbed dynamics. At equilibrium,
this relation is remarkably simple, still very deep, as it only
involves the correlation between the observable considered and the
quantity which is coupled with the perturbing field.  Out of
equilibrium, it is still possible to relate the response function to
unperturbed correlators, but other terms have to be taken into
account, which explicitly depend on the dynamics of the model. In
recent years many generalizations have been derived (see for
instance~\cite{CKP94,LCZ05,ss06,LCSZ08,BBMW10}).

\begin{figure}[t!]
\includegraphics[width=1.\columnwidth,clip=true]{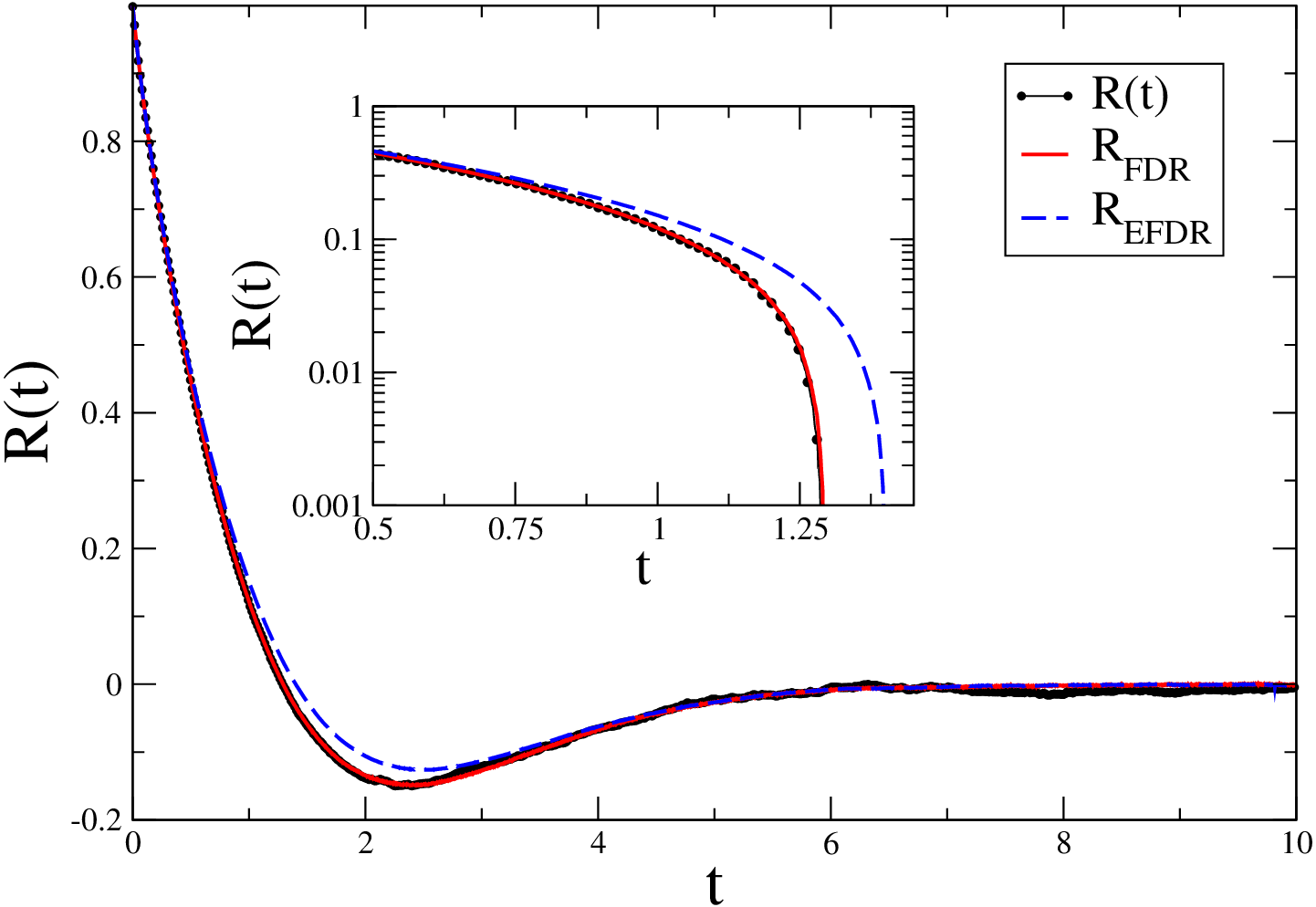}
\caption{(Color online) Response functions measured in numerical
  simulations of the model with Coulomb
  friction~(\ref{coulomb}). Parameters are $\gamma=0.05$, $\Delta=1$,
  $\gamma T=0.5$, $\mu = 0.25$ and perturbation $h=0.1$. Inset: zoom
  of the discrepancy region in semi-log scale.}
\label{fig_FDR_coul} 
\end{figure}

\begin{figure}[t!]
\includegraphics[width=1.\columnwidth,clip=true]{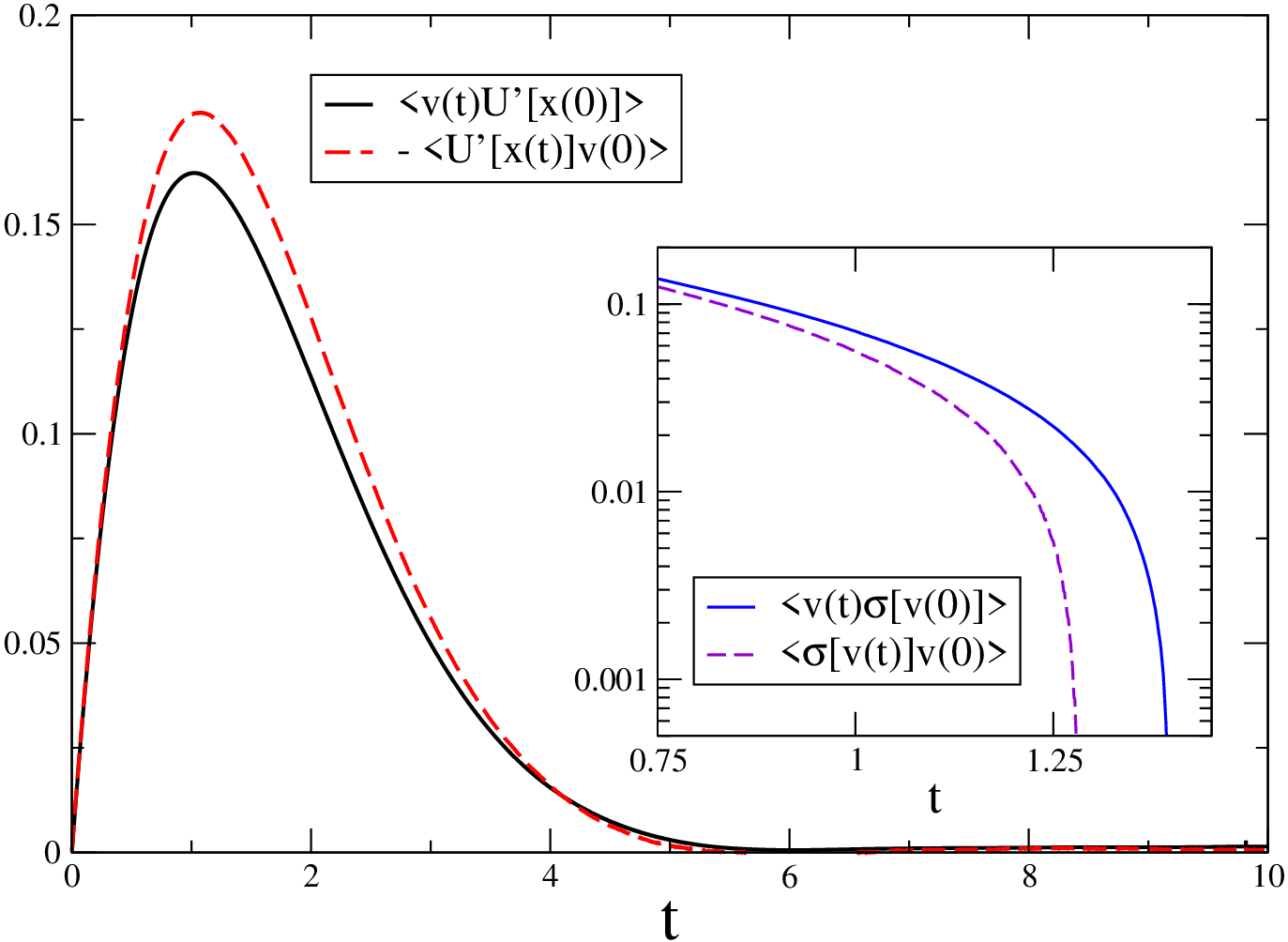}
\caption{(Color online) Comparison among the correlation functions
  appearing in Eq.~(\ref{resp}) ($\langle v(t)U'[x(0)]\rangle$ and
  $\langle U'[x(t)]v(0)\rangle$ in the main frame) measured in
  numerical simulations with parameters: $\gamma=0.05$, $\Delta=1$,
  $\gamma T=0.5$ $\mu = 0.25$.  Inset: Comparison among the
  correlation functions $\langle v(t)\sigma[v(0)]\rangle$ and
  $\langle\sigma[v(t)]v(0)\rangle$ in semi-log scale.}
\label{fig_FDR_coul2} 
\end{figure}

Here we are interested in the use of the FDR as a tool to point out
nonequilibrium conditions.  Therefore we investigate the FDR for the
model introduced above in the stationary regime.  We start by deriving
a general expression for the response function $R(t,t')$ of the velocity $v(t)$
to a perturbation $h(t')$ applied at a previous time $t'$:
\begin{equation}
R(t,t')=\left .\frac{\delta \langle v(t)\rangle}{\delta
  h(t')}\right|_{h=0}.
\label{fdr.00}
\end{equation}
Since for Langevin-like systems with Gaussian noise the response can
be expressed as~\cite{KV94}
\begin{equation}
R(t,t')=\frac{1}{2\gamma T}\langle v(t)\eta(t')\rangle,
\label{fdr.0}
\end{equation}
substituting the expression for the noise, and considering a
stationary state, where time-translational invariance holds, fixing $t'=0$ we can write
\begin{equation}
R(t)=\frac{1}{2\gamma T}\Big\{\langle v(t)\dot{v}(0)\rangle 
+\langle v(t)F[v(0)]\rangle
+\langle v(t)U'[x(0)]\rangle\Big\}.
\label{resp1.0}
\end{equation}
Then, exploiting the fact that $\langle v(t)\dot{v}(0)\rangle=-\langle
\dot{v}(t)v(0)\rangle$, and that $\langle \eta(t)v(0)\rangle=0$ by
causality, we can recast Eq.~(\ref{resp1.0}) in the form
\begin{eqnarray}
R(t)&=&\frac{1}{2\gamma T}\Big\{\langle v(t)F[v(0)]\rangle 
+\langle F[v(t)]v(0)\rangle  \nonumber \\
&+&\langle v(t)U'[x(0)]\rangle+
\langle U'[x(t)]v(0)\rangle\Big\}.
\label{resp.0}
\end{eqnarray}
This expression represents an extension to inertial cases of the result
obtained in~\cite{CKP94} for overdamped dynamics.  Notice that in the case of a quadratic
potential $U(x)=kx^2/2$, Eq.~(\ref{resp.0}) can be further simplified,
exploiting the relation $\langle x(t)v(0)\rangle=-\langle
v(t)x(0)\rangle$, and one gets
\begin{equation}
R(t)=\frac{1}{2\gamma T}\left\{\langle v(t)F[v(0)]\rangle 
+\langle F[v(t)]v(0)\rangle\right\}.
\end{equation}
When an \emph{equilibrium} stationary state is attained, using the
Onsager reciprocity relations $\langle v(t)F[v(0)]\rangle= \langle
F[v(t)]v(0)\rangle$ and $\langle v(t)U'[x(0)]\rangle=-\langle
U'[x(t)]v(0)\rangle$, one finds the EFDR
\begin{equation}
R_{EFDR}(t)=\frac{1}{\gamma T}\langle v(t)F[v(0)]\rangle.
\label{resp2.0}
\end{equation}

In the presence of Coulomb friction~(\ref{coulomb}),
Eq.~(\ref{resp.0}) takes the following expression, that we denote by
$R_{FDR}$
\begin{eqnarray}
R_{FDR}(t)&=&\frac{1}{T}\Big\{\langle v(t)v(0)\rangle \nonumber \\
&+&\frac{\Delta}{2\gamma}\left[\langle v(t)\sigma[v(0)]\rangle 
+\langle\sigma[v(t)]v(0)\rangle \right] \nonumber \\
&+&\frac{1}{2\gamma}\left[\langle v(t)U'[x(0)]\rangle+
\langle U'[x(t)]v(0)\rangle\right]
\Big\}.
\label{resp}
\end{eqnarray}
As shown in Sec.~\ref{DB}, in the absence of external potential an
equilibrium state is attained and one obtains the EFDR with Coulomb
friction
\begin{equation}
R_{EFDR}(t)=\frac{1}{T}\left\{\langle v(t)v(0)\rangle
+\frac{\Delta}{\gamma}\langle v(t)\sigma[v(0)]\rangle \right\}.
\label{resp2}
\end{equation}

In Fig.~\ref{fig_FDR_coul} we report numerical simulations of the
response function $R(t)$ for the model with dry
friction~(\ref{coulomb}), in the presence of the ratchet potential,
Eq.~(\ref{ratchet_pot}), and we compare it with the EFDR,
Eq.~(\ref{resp2}), and the generalized FDR, Eq.~(\ref{resp}):
time-reversal symmetry is clearly broken in this system and
nonequilibrium contributions have to be taken into account to obtain
the correct expression for the response function.  Numerical
simulations are performed by integrating the Kramers equation with a
time step $\delta t=10^{-5}$. The response is measured as the
difference between the mean velocity in the presence of a small
perturbation $h=0.1$ and the mean velocity for zero field, divided by
$h$.  The linear regime has been first checked by applying different
values of perturbation.  Data are averaged over about $10^6$
realizations. Notice that the response function shows a non-monotonous
behavior, taking negative values. This is a phenomenon typical of
inertial dynamics, and in our case it is very pronounced, at variance
with the behavior reported in~\cite{CC12} for a flashing ratchet
model, where, for the chosen parameters, it is barely visible.

In Fig.~\ref{fig_FDR_coul2} we also report the correlation functions
appearing in Eq.~(\ref{resp}) to explicitly show the violation of
time-reversal symmetry in this case.  Analogous results (not shown),
with non-monotonic, oscillatory behaviors for the response functions,
have been obtained using the frictional force of the model for active
particles, Eq.~(\ref{active}).

\section{Ratchet effect}
\label{rat}

In this Section we show how the presence of a spatial asymmetry
coupled with nonequilibrium conditions produces a net average motion
of the particle described by the model in Eq.~(\ref{kramers}), with
different frictional forces.  In particular, here we investigate the
ratchet effect driven by Coulomb friction, recently shown in different
systems~\cite{gnoli,SGP13}, in the context of Langevin equations.
Notice that, at variance with the previous models of Brownian motors
studied in the literature~\cite{R02,HM03}, such as flashing ratchets,
rocking ratchets, and deterministically forced ratchets, in our model
there are no external forces driving the motor, and the only source of
dissipation is the presence of Coulomb friction.

\begin{figure}[t!]
\includegraphics[width=1.0\columnwidth,clip=true]{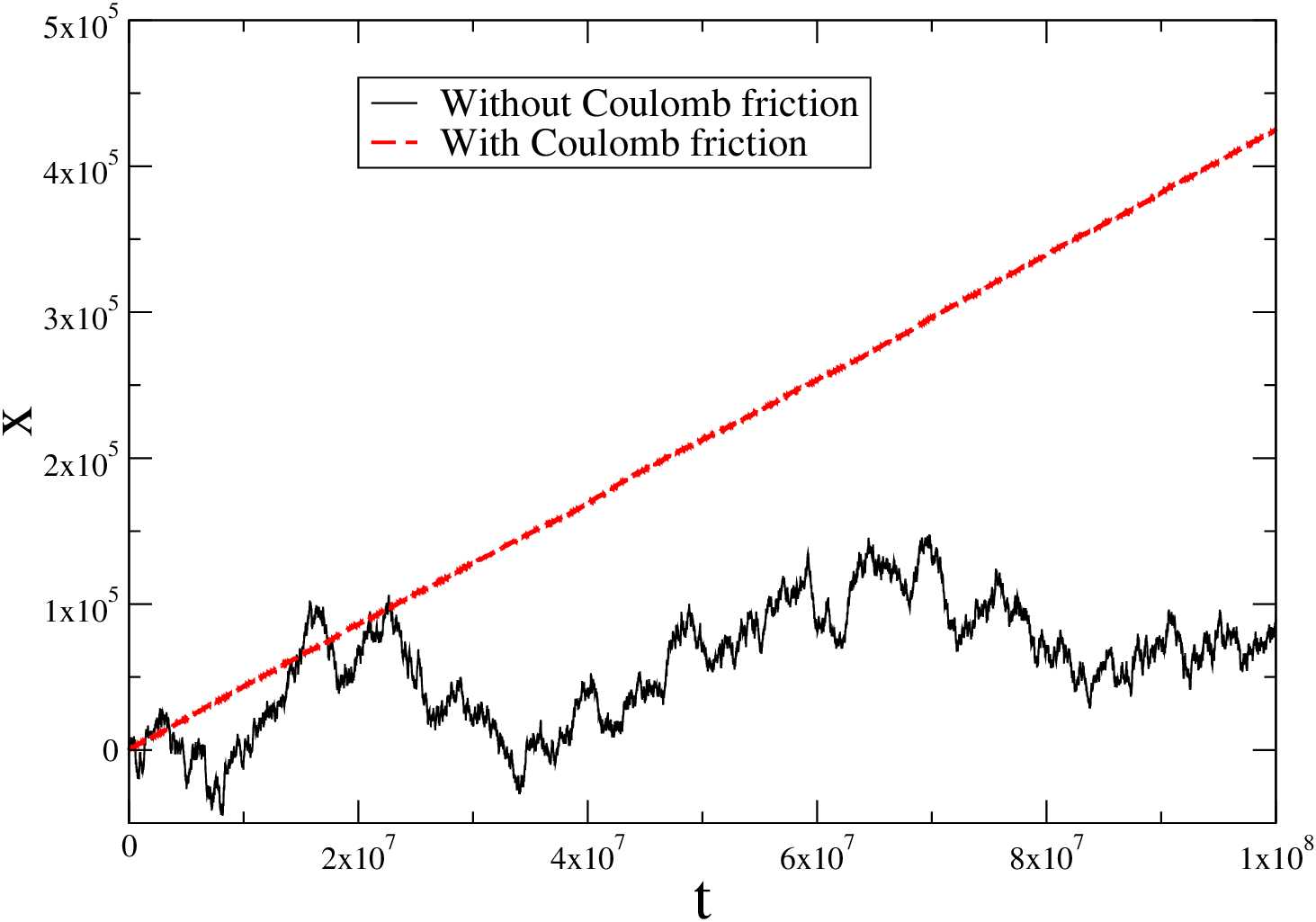}
\caption{(Color online) Time evolution of the position $x(t)$ for the
  model in Eq.~(\ref{kramers}) without Coulomb friction (black
  continuous lines), with parameters $\gamma=0.05$, $\gamma T=0.5$,
  and $\mu=0.4$, and with Coulomb friction (red dashed line), with
  same parameters and $\Delta=1$.}
\label{traj} 
\end{figure}

\begin{figure}[t!]
\includegraphics[width=1.0\columnwidth,clip=true]{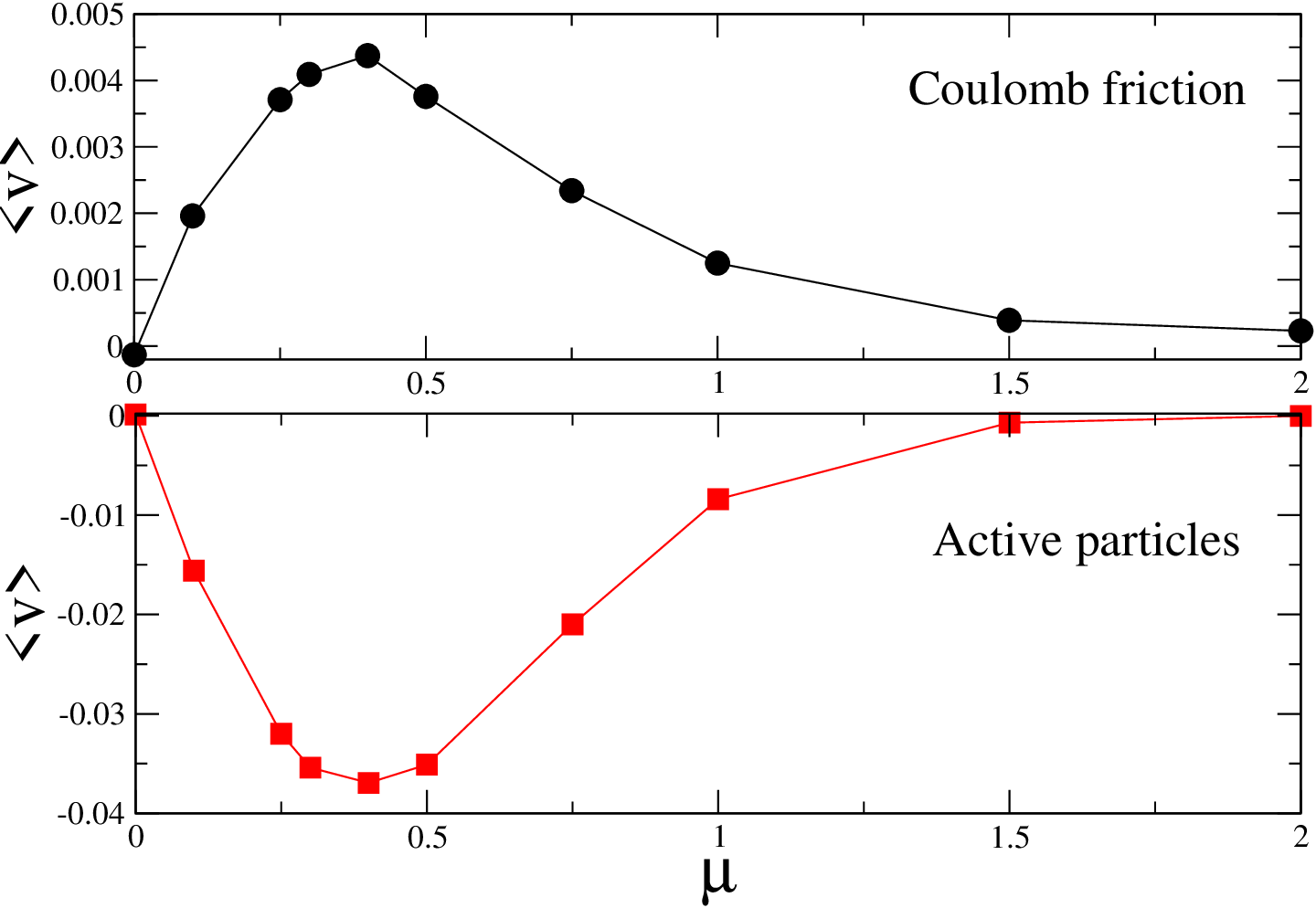}
\caption{(Color online) Average drift for the model with Coulomb
  friction~(\ref{coulomb}) (top panel) with parameters $\Delta=1$,
  $\gamma=0.05$, $\gamma T=0.5$, and for the model for active
  particles~(\ref{active}) (low panel) with parameters
  $\gamma_1=\gamma_2=1$, and $\gamma T=0.5$, as a function of the
  asymmetry parameter $\mu$ of the spatial potential,
  Eq.~(\ref{ratchet_pot}).}
\label{drift_mu} 
\end{figure}

In Fig.~\ref{traj} we show the time evolution of the position of the
Brownian particle described by Eq.~(\ref{kramers}), in the absence
(continuous black line) and in the presence of Coulomb friction
(dashed red line). Only in the latter case a net ratchet effect can be
clearly observed, whereas in the former case one observes large
fluctuations around zero. This is due to the fact that, as discussed
in Sec.~\ref{DB}, if only linear friction is considered, the system
attains an equilibrium stationary state and no ratchet effect can
occur.  On the other hand, it is interesting to study the behavior of
the system upon varying the parameters entering the model. In
particular, in Fig.~\ref{drift_mu} (top panel) we report the average
velocity of the particle described by Eq.~(\ref{kramers}) with Coulomb
friction~(\ref{coulomb}), as a function of the asymmetry parameter
$\mu$ of the ratchet potential, Eq.~(\ref{ratchet_pot}). In the lower
panel of the same figure, we also report the results of numerical
simulations of the model for active particles described by
Eq.~(\ref{active}). As expected, for $\mu=0$, the ratchet effect
vanishes in both models, because the potential is spatially symmetric
in that case. By increasing the value of $\mu$ a non-monotonic
behavior is observed: notice, in particular, that a peak at the same
value $\mu\simeq 0.4$ is observed for both the models (see
Fig.~\ref{drift_mu}). This means that the maximization of the ratchet
effect as a function of the shape of the potential is independent of
the specific models considered. The decreasing of the ratchet effect
for large values of $\mu$ is probably due to the fact that the
potential develops more than one minimum. This causes an overall
slowing down of the dynamics and, therefore, of the average velocity.

\section{Conclusions}
\label{concl}

In this article we have presented a numerical study of the Kramers
equation describing the motion of a particle in an asymmetric
potential and in the presence of nonlinear frictional forces, in
particular the Coulomb frictional force.  We have characterized the
nonequilibrium behavior of the system by studying the generalized
fluctuation-dissipation relation, pointing out the breaking of
time-reversal symmetry when both nonlinear velocity-dependent friction
forces and external potential are present. In such situations, even if
no other external forces are present and the only source of
fluctuations is the thermal noise of the Langevin equation, we have
shown that the Coulomb friction can be a source of dissipation
sufficient to drive a motor effect.
 
This kind of ratchet effect could be studied in experiments in nano-
and micro-devices, where the presence of dry friction is still present
and strongly affect the dynamical properties of the
system~\cite{GTVT10,VMUZT13}.

It would be also very useful to obtain analytical approximate
expressions for the average drift of the particle in the presence of
Coulomb friction.  Moreover, the ratchet effect should be further
characterized by studying its energetics and efficiency at maximum
power, when also an external load is applied to the system.

\begin{acknowledgments} The author acknowledges the kind hospitality of the KITPC
at the Chinese Academy of Sciences, during the Program ``Small system
nonequilibrium fluctuations, dynamics and stochastics, and anomalous
behavior'', when this article was completed.  He also thanks
A.~Puglisi, H.~Touchette, A.~Imparato, and J.~Talbot for useful
discussions.  The work of the author is supported by the
``Granular-Chaos'' project, funded by the Italian MIUR under the
FIRB-IDEAS grant number RBID08Z9JE.
\end{acknowledgments}

\bibliography{fluct.bib}

\end{document}